\documentstyle[preprint,aps,prd]{revtex}
\begin{document}  
\input{psfig}
%\draft                                                               
\title{A Study of Cosmic Ray Composition in the Knee
                  Region using Multiple Muon Events in the
                  Soudan 2 Detector}
\author{S.M. Kasahara$^b$, W.W.M. Allison$^c$, G.J. Alner$^d$, D.S. Ayres$^a$, W.L. Barrett$^f$, 
C.R. Bode$^b$,\\
P.M. Border$^b$, C.B. Brooks$^c$, J.H. Cobb$^c$, D.J.A. Cockerill$^d$, 
R.J. Cotton$^d$, H. Courant$^b$,\\ 
D.M. DeMuth$^b$, B. Ewen$^e$, T.H. Fields$^a$, H.R. Gallagher$^c$,
M.C. Goodman$^a$, R.W.Gran$^b$, R.N. Gray$^b$, 
K. Johns$^{b,}$\footnote{Now at University of Arizona, Physics Department, Tucson, AZ 85721, USA},
T. Kafka$^e$, W. Leeson$^e$, P.J. Litchfield$^d$,
N.P. Longley$^{b,}$\footnote{Now at Swarthmore College, Swarthmore, PA 19081, USA},\\
M.J. Lowe$^{b,}$\footnote{Now at Dept of Medical Physics, University of Wisconsin, Madison, WI 53705, USA},
W.A.  Mann$^e$, M.L. Marshak$^b$, E.N. May$^a$, R.H. Milburn$^e$, W.H. Miller$^b$,\\
L. Mualem$^b$, A. Napier$^e$, W. Oliver$^e$, G.F. Pearce$^d$, E.A. Peterson$^b$, L.E. Price$^a$,\\
D.M. Roback$^{b,}$\footnote{Now at Dept of Radiology, University of MN, Minneapolis, MN 55455, USA},
K. Ruddick$^b$,
D.J. Schmid$^{b,}$\footnote{Now at Kodak Health Imaging Systems, Dallas, TX, USA},
J. Schneps$^e$, M.H. Schub$^b$,
R.V. Seidlein$^a$,\\
M.A. Shupe$^{b,*}$, N. Sundaralingam$^{e,}$\footnote{Now at Edward Waters College, Jacksonville, FL 32209, USA},
J.L. Thron$^a$, H.J. Trost$^{a,}$\footnote{Now at MicroFab Technologies, Plano, TX, USA}, J.L. Uretsky$^a$,\\ 
V. Vassiliev$^b$, G. Villaume$^b$, S.P. Wakely$^b$, D. Wall$^e$,
S.J. Werkema$^{b,}$\footnote{Now at Fermi National Accelerator Laboratory, Batavia, IL 60510, USA}
and N. West$^c$}
\address{$^a$ Argonne National Laboratory, Argonne IL 60439, USA;\\
         $^b$ University of Minnesota, Minneapolis MN 55455, USA;\\
         $^c$ Department of Physics, University of Oxford, Oxford OX1 3RH, UK;\\
         $^d$ Rutherford Appleton Laboratory, Chilton, Didcot, Oxfordshire, OX11 0QX, UK;\\
         $^e$ Tufts University, Medford MA 02155, USA;\\
         $^f$ Western Washington University, Bellingham WA 98225, USA}
\date{\today}
\maketitle
\begin{abstract}        
Deep underground muon events recorded by the Soudan 2 detector, located
at a depth of 2100 meters of water equivalent, have been
used to infer the nuclear composition of cosmic rays in the
``knee'' region of the cosmic ray energy spectrum.  The observed
muon multiplicity distribution favors a composition model with
a substantial proton content in the energy region $8\times 10^5-1.3\times10^7$ GeV/nucleus.
\end{abstract}
% body of paper here     
%\narrowtext    
\section{Introduction}\label{sec:intro}
\par  The composition of the cosmic rays in the ``knee'' 
($\sim 10^4$ TeV/nucleus) region of the cosmic ray all-particle spectrum has 
consequences for astrophysical models of particle acceleration and 
propagation.  
For example, a model of accretion onto a black
hole in the center of an Active Galactic Nucleus predicts an excess of
protons around the knee region in the cosmic ray flux \cite{agn4},
while a model of shock acceleration during a
supernova blast into a stellar wind environment predicts
an excess of heavy nuclei in the cosmic ray flux in the same energy
region \cite{bier1}. 
\par Unfortunately, the composition at energies 
above $\sim 1000$ TeV/nucleus is
difficult to measure directly due to the steeply falling spectrum
of cosmic ray primaries.  The flux of particles with energies
greater than 1000 TeV/nucleus is only $\sim$ 60/m$^2$/sr/year.
Therefore measurements of cosmic rays in this energy regime require
detectors of either very large acceptance or long exposure, neither
of which is currently feasible for measurements near
the top of the atmosphere or above.   
Instead, studies of the composition of cosmic ray primaries in this
high energy regime can be carried out indirectly through measurements
of aspects of the atmospheric cascade generated by the
interaction of a cosmic ray primary with the earth's atmosphere. 
\par Deep underground experiments, such as Soudan 2, indirectly study the 
composition of cosmic rays by comparing
observations of multiple muon event rates to expectations
derived through Monte Carlo simulations using various trial composition 
models as input.  A multiple muon
event is one in which two or more nearly parallel, time-coincident
muon tracks are observed in the detector.  These muons are decay products
of mesons which are generated in the hadronic core of
the atmospheric cascade.
At high energies, massive primaries generate more high energy muons
per event than proton primaries of the same total energy.  
This is
because the initial parent meson (predominantly pion) particle multiplicities are larger
from the more massive primary, and the atmosphere is more 
favorable to pion decay versus interaction early in the cascade development.
In addition, the point of first interaction of the heavy primary is
likely to be higher in the atmosphere than for the proton primary due to the larger cross-section
of the heavy nucleus-air interactions.
\par At the Soudan site, the measurement of the high energy
muon component underground is coupled with sampling of the
electromagnetic component of the air shower at the surface
using a small proportional tube array \cite{nat}, and a Cherenkov light
detector array \cite{cheren}.  A correct interpretation of the cosmic ray
composition should yield consistent results using all three experimental 
techniques in all possible combinations.        
\par In this paper, we report on an analysis in which the
observed multiple muon event rates recorded in the Soudan 2 detector
are compared with the simulated rates obtained using three distinct
trial composition models.  The next section of this paper contains a brief 
description of the relevant aspects of the detector. 
In Sec.\ \ref{sec:data}, we report on the analysis of the observed
multiple muon event rates.  Sec.\ \ref{sec:mc} contains a discussion
of the Monte Carlo simulation.  Sec.\ \ref{sec:analysis}
has a discussion of the test composition models used in this analysis
and a comparison between the data multiple muon rates and the rates
observed using these models.   Finally, 
Sec.\ \ref{sec:conclude} summarizes the results.

\section{The Detector}\label{sec:detector}
The Soudan 2 detector \cite{dco} is a high resolution tracking 
calorimeter located in the
Soudan Underground Mine State Park in northern Minnesota,
at a depth of 710 meters below the surface of the earth.  This
depth corresponds to a muon threshold energy of $\sim 0.7$ TeV for
a muon transmission probability of at least 50$\%$.
The modular design of the detector has allowed the
continuous acquisition of data from the beginning of its
construction in mid-1988 to the present, during which time Soudan 2
has recorded more than 33 million muon events.  The detector reached 
its full size of 224 1 m$\times$1 m$\times$2.7 m modules in November,
1993, for a final operating size of 8 m$\times$14 m in horizontal
surface area $\times$5.4 m in height.
\par The active region of each detector module consists of 
7560 15 mm diameter plastic drift tubes layered between
241 1.6 mm thick corrugated steel sheets. Ionization deposited
in the tubes drifts under the influence of an electric
field toward the tube ends, where it is collected on
vertical anode wires spaced 15 mm apart,
and horizontal cathode strips spaced 10 mm apart.   
The pulses on the wires and strips are read-out and
digitized every 200 ns, which, for a drift rate of 0.6 cm/$\mu$s,
corresponds to a timing resolution along the drift axis of 1.2 mm.  (The 
measured resolution along the drift-axis of $\sim 6$ mm is larger than 
the timing resolution due to variations in drift velocities.)
Pulse matching of the anode and cathode signals yields a 
three-dimensional space coordinate for each drift tube crossing along
the path of a charged particle.
\par
\begin{figure}
\centerline{\psfig{figure=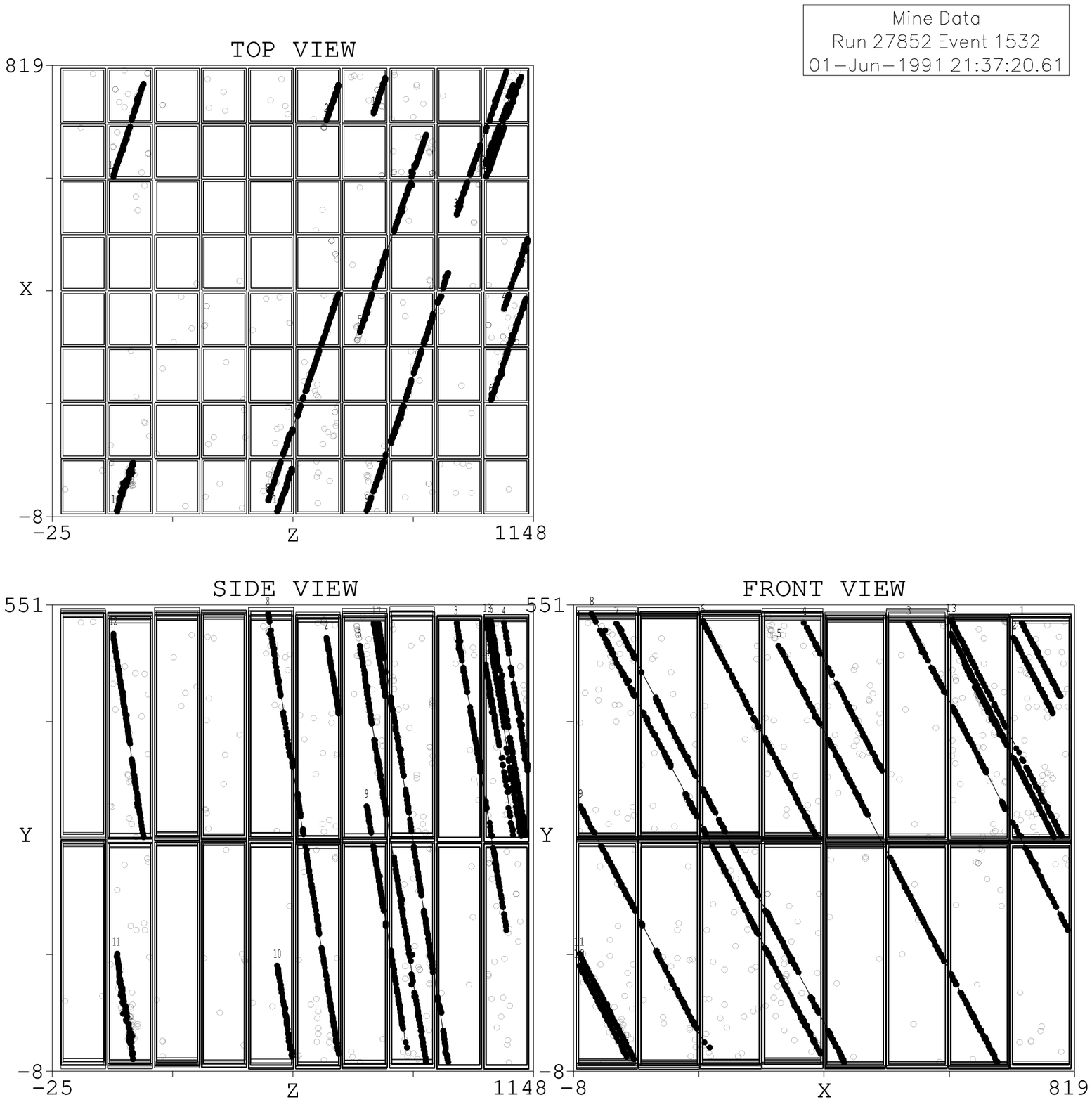,height=6.in,bbllx=0bp,bblly=0bp,bburx=600bp,bbury=650bp,clip=}}
\caption[]{\label{fig:ev14}
A 14 muon event in the Soudan 2 detector.  The figure shows three
different view of the same event.  All axes are labeled in centimeters.
}
\end{figure}

 Events with high muon multiplicity require the ability to separate 
tracks bunched tightly together, as well as to distinguish tracks 
obscured by showers in multiple muon events in which a large
bremsstrahlung has occured.  The high resolution of Soudan 2
is particularly suited to this type of study since a typical
track will leave hundreds of reconstructed pulses along its
path.  Fig.~\ref{fig:ev14} shows three views of a 14-muon event 
as recorded in the detector.  The resolution of a space point
along a muon track detected in Soudan 2 is $\sim 1$ cm.  The
angular resolution of a typical muon track is $< 1^o$.          
\par
\begin{figure}
\centerline{\psfig{figure=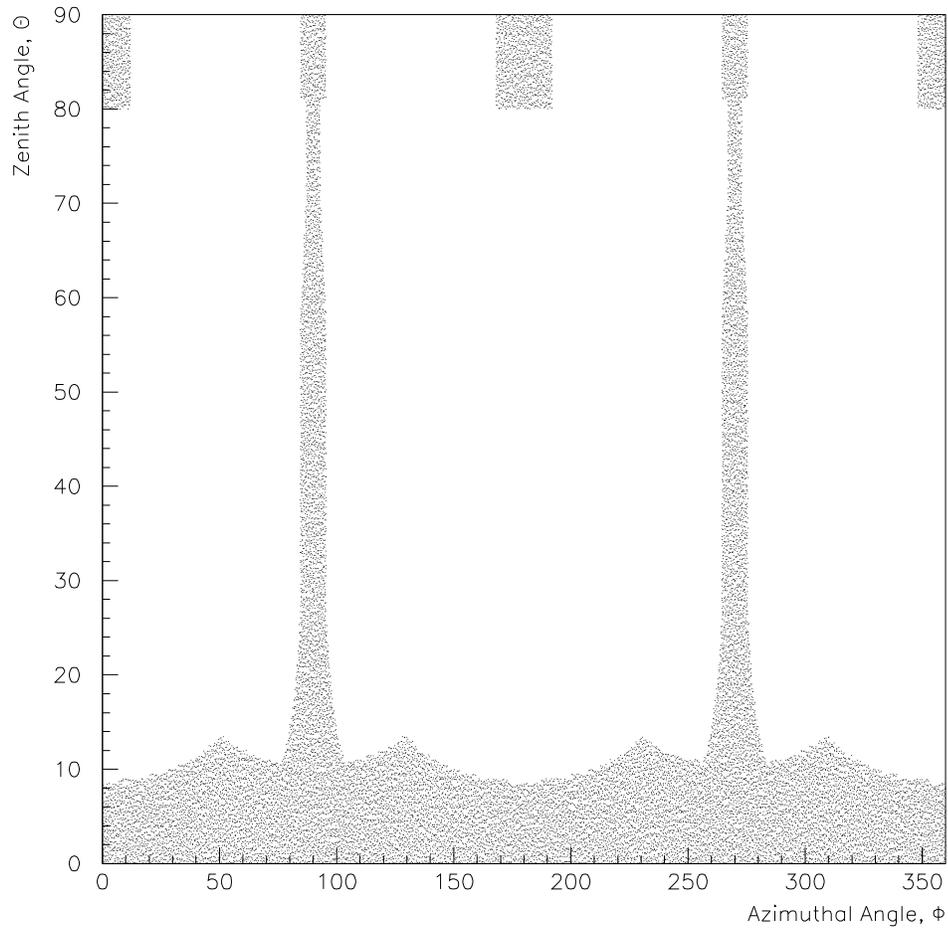,angle=-90.,height=6.in,bbllx=0bp,bblly=100bp,bburx=600bp,bbury=650bp,clip=}}
\caption[]{\label{fig:trigger} 
The shaded regions correspond to the ``idealized'' trigger holes of the detector as seen by muons with a track
length of 50 cm.  The coordinate system is defined such that ($\phi=0^o$,$\theta=90^o$) points North, 
($\phi=90^o$,$\theta=90^o$) points West, and $\theta=0^o$ points straight up from
the detector floor plane.}
\end{figure}

 The primary trigger operating over the time span of this
data sample was called the ``edge'' trigger.  It required a muon
to have pulsed a minimum of 7 anode wires or 8 cathode strips 
out of any contiguous block of 16 channels of either type, separated 
by at least one trigger clock pulse of length 600 ns, and occuring 
within the given time window of 72$\mu$s.  The effect of the trigger 
requirement on a muon track of length 50 cm, the minimum track length 
considered in this analysis, is shown in Fig.~\ref{fig:trigger}.  The
boundary shown in the figure represents the ideal; in actuality, 
muon bremsstrahlung or pair production initiated showers assist
triggers allowing some muons outside of this region to satisfy
the trigger requirements as well.  This, combined with individual channel
inefficiencies, creates some fuzziness at the trigger
acceptance boundaries.  To avoid the trigger holes, the data events 
in this analysis were subjected to software imposed zenith and
azimuthal angle cuts as discussed in the next section.
\par In addition to the main detector, there is a 14m$\times$31m$\times$10m veto 
shield consisting of proportional tubes which nearly covers 
the entire $4\pi$ steradian surrounding the main detector.  
For multiple muon analysis, the shield is 
useful as a tool for selecting multiple muon event candidates in the 
main detector.  It can also be used to study muon tracks which
pass outside of the main detector volume \cite{sundar},
however this has not been included in the analysis presented
here.  
\section{Data Acquisition and Analysis}\label{sec:data}    
The data set reported in this paper consists of a sample of the 
total number of events recorded in Soudan 2 
over the time span June 1991 through October 1991, and
includes 7.2$\times 10^5$ single muon and 22000 multiple muon 
events after all cuts are applied. The detector size at the time 
of this data sample was 8 m$\times$11 m in surface area $\times$5.4 m high.  
A larger data sample is not necessary for this analysis, since 
its accuracy is dominated by systematic uncertainties in the Monte
Carlo simulation and not by statistics.
\par Data in Soudan 2 are accumulated in modular 
``runs'' each lasting a little over an hour and containing
typically 1500-2500 events.  To be
included as part of the data sample, each run needed to pass
a series of ``run quality'' checks in order to exclude runs
with localized hardware failures such as high voltage trips, 
excessive noise, or anomalously high or low average trigger rates.     
After application of run cuts, the total live-time
considered in this analysis was 1348.8 hours.
\par Several cuts were applied to the data sample to ensure
the highest quality data.  Each muon track was 
required to have a minimum length of 50 cm.  
Multiple muon events were required to satisfy a ``parallel''
cut such that for each muon track there was an angular separation of 
less than $5^o$ from at least one other muon in the group.  This
restriction was to eliminate locally produced tracks.
To avoid the trigger holes, we have made cuts on the azimuthal
and zenith angle regions of acceptance.  The muon events
are confined to the azimuthal regions $10^o<\phi<80^o$ in
each quadrant and zenith angles $>15^o$.  The muon events
are further confined to zenith angles $<60^o$ as this allows
us to apply the flat atmosphere model which is
prevalent in atmospheric cascade simulations.  
The zenith and azimuthal angles of an event with more than one muon
track were determined as the average of all muon tracks satisfying
the length and parallel cuts.          
\par By design, the Soudan 2 track reconstruction software
finds and reconstructs everything from through-going muons
to ``contained'' tracks consisting of as few
as 5 pulses.  This is because Soudan 2's primary purpose is 
to search for the short tracks left by proton decay.  The software is
very efficient at finding and reconstructing single
muon events which typically contain several hundred
pulses.  
For this reason, determining the single
muon event rate required only purifying the total
of all tracks found by the reconstruction software to extract
the through-going muon tracks.
The regular track reconstruction code was modified 
to tag muon tracks as those which satisfied at least {\it one} of the following
criteria on {\it both} ends of the track:        
\begin{itemize}  
\item{The track extrapolated to a time-coincident veto shield hit.}
\item{The last reconstructed hit on the end of the track was
      within 50 cm of the detector edge.}
\item{The end of the track projected through a detector crack,
      which is defined as the small open region between each
      pair of modules.}
\end{itemize}          
These requirements were enough to make the software very efficient
at producing a very pure single muon sample. 
The single muon event reconstruction was tested against 1900 randomly
selected events from two separate data runs,
of which 594 events were found to be single muon
events passing the length and angular cuts considered here.   
The results of this test are shown in Table \ref{tab:eff}.  
The software was found to be $99.2\pm0.4\%$ efficient at
identifying single muon events which passed the stated angular
and length cuts.  The background of misreconstructed tracks
contributing to the single muon sample was determined to be $2.0\pm0.6\%$.
These corrections
have been applied to the observed number of single muon events in Table \ref{tab:mult}.                         
\par The software reconstruction
of multiple muon events in the main detector is complicated due to
a hardware design feature which electrically adds signals from
separate regions of the detector together during readout.
The purpose of this ``multiplexing'' is to 
decrease the cost of the electronics
required to read out the $4\times10^6$ drift tubes, and it   
occurs just before the pulses are digitized.
The multiplexing is configured
so that a match between a given anode pulse    
and cathode pulse has a unique position in the detector.
The pulses are demultiplexed at the software
stage.  This multiplexing has
little effect on the software reconstruction of contained events
or single muon events because the relatively small
number of pulses in these events leads to a simple interpretation
of the data, but as the multiplicity of the event
increases, the number of pulses and the complexity of the demultiplexing of the
event increases as well.  For this reason, the software
reconstruction of all candidate multiple muon events was 
supplemented by scanning performed by a physicist 
using an interactive graphics program.    
\par Multiple muon candidates were selected based on a set
of generous but simple criteria.  These criteria were that a
candidate event contain both of the following:
\begin{itemize}
\item{At least one ``good'' track satisfying the
      criteria of the single muon tracks described above, except that in
      this case the angular restrictions on this one track
      were loosened to $12.5^o \leq \theta \leq 62.5^o$ and
      $7.5^o \leq \phi \leq 82.5^o$ in each quadrant.} 
\item{At least one 2-dimensional anode-time or
      cathode-time reconstructed track, which, when paired with 
      the opposite
      2-d track from the ``good'' track, was parallel to the
      ``good'' track within $5^o$.  The sum of the projected lengths
      of all parallel 2-d tracks in either the anode-time or
      cathode-time view had to be at least 50 cm.}
\end{itemize}    
The net effect of these cuts was to select $\sim 6\%$
of the total number of muon events as multiple muon candidates,
while the final sample post-scanning consisted of only
$3\%$ multiple muons.  Therefore these selection cuts are considered
to be very liberal.  
\par The multiple muon candidates were examined not
just for multiplicity but to check their reconstruction results
which were displayed directly over the event.  The 
majority of low multiplicity events ($N_\mu < 4$) had been
reconstructed correctly by the software, while the highest
multiplicity events ($> 8$) generally required some manual corrections
which were applied through the interactive graphics program
to produce the correct fits.               
\par Events with multiplicities greater than 12 were discarded
in this analysis due to a flaw in the data acquisition hardware which existed at
the time of the data sample.  This flaw put a maximum limit
on the amount of readout time taken by any one event from
the time of the trigger to the end of the event readout CAMAC sequence.
The amount of time spent in this sequence rises with the
size of the event, which is in turn dependent on the event
multiplicity.  Since the readout time distributions corresponding to a
given multiplicity have long tails, there is a gradual lessening
in the efficiency to readout the entire event with increasing multiplicity.
Through a simulation in which low multiplicity data events (for which the
readout time distributions are known) were used to generate the
readout time distributions of high multiplicity events, 
the efficiency for reading out high multiplicity events was determined \cite{thesis}.  
These efficiencies were calculated based on a worst case scenario, and as 
such cannot be used to correct the data.  In this simulation,
99\% of 12 muon events were read out successfully by the
data acquisition software.  This is the maximum multiplicity
considered in this analysis.
\par The efficiency of the multiple muon candidate selection criteria
is tabulated in Table \ref{tab:mmueff}.  In this case,
the same two runs used to determine the single
muon efficiency were used to determine the number of
multiple muon events which satisfy the angular, length,
and parallel cuts described above.  Of the 2970 events
in these two runs scanned for multiple muon events, 30 events were determined to be
multiple muon events which satisfied these cuts, and the multiple
muon selection software found all 30 of these events.  Five
additional runs shown in the table were scanned by a team
of undergraduate scanners.  Seventy-nine out of the 8830  events
scanned were found to be multiple muon events which
satisfied the cuts, of which the multiple muon selection
software found 79.  The 
efficiency of the multiple muon selection software has therefore 
been determined in excess of $97.5\%$ at the $90\%$ confidence level.  
This uncertainty has been applied to the errors
in Table \ref{tab:mult} for all multiplicities,
even though the selection software will certainly be more
efficient at finding high multiplicity events than
low multiplicity events, so that this can be considered
to be a very conservative estimate of the total uncertainty
at high multiplicities.
\par
\begin{figure}
\centerline{\psfig{figure=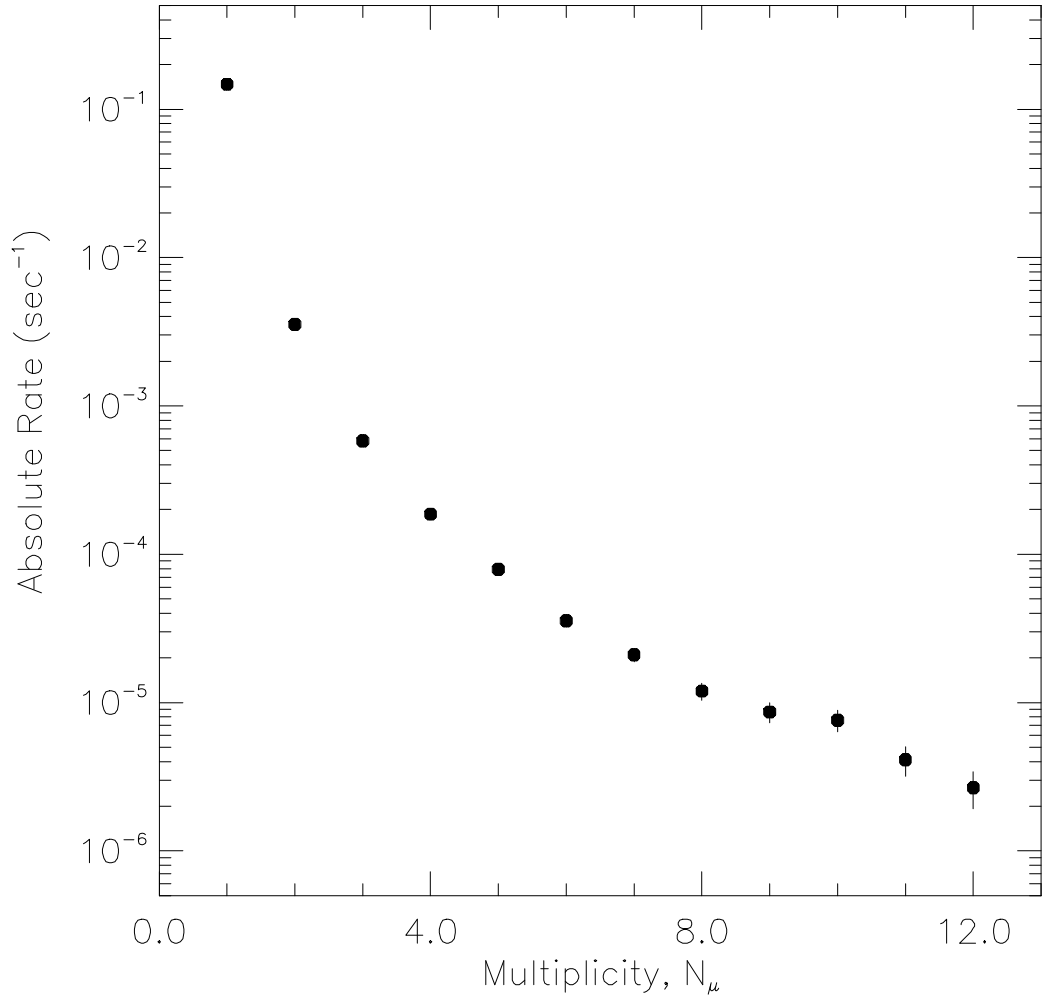,height=6.in,bbllx=0bp,bblly=100bp,bburx=600bp,bbury=650bp,clip=}}
\caption[]{\label{fig:abs} 
The absolute event rate as a function of multiplicity as seen in Soudan 2.  
The rates are also tabulated in Table~\ref{tab:mult}.
}
\end{figure}

 The final corrected muon event rates observed in Soudan 2
are shown in Table \ref{tab:mult} and Fig.~\ref{fig:abs}. 

\section{Monte Carlo Simulation}\label{sec:mc}                  
\par The increasing statistical accuracy of indirect measurement 
data correlated to phenomena from primary energies in the knee region 
requires the use of sophisticated Monte Carlo simulations
to constrain the systematic errors in these types of measurements.
In this application, we have used the most fully-developed 
Monte Carlo simulation of the atmospheric cascade and
muon rock propagation currently available. 
The Monte Carlo simulation used in this analysis consists 
of three stages:
\begin{itemize}
\item A 3-dimensional simulation of the atmospheric
      cascade.

\item A 3-dimensional simulation of the propagation
      of the muons through the rock overburden.

\item A simulation of the detector.
\end{itemize}
\par The cascade simulation uses the HEMAS cascade code \cite{hemas} 
for controlling the overall structure
of the cascade development.  This code injects a nucleus 
of a requested energy, mass, azimuth and zenith angle
into the atmosphere.  It then tracks this particle and any
secondaries produced along the path of the cascade 
development until they either interact, decay, drop
below some user defined energy threshold, or 
reach the atmospheric sampling height.  In our case,
this last quantity is at the surface of the earth
above Soudan 2, which corresponds to 492 meters 
above sea level.
\par The HEMAS cascade code is built so that 
the hadronic interaction code is fully modular.  In this 
analysis, we have used
the recent program SIBYLL \cite{sibyll} for generating
hadronic interactions.  SIBYLL is based on the dual parton model with
minijet production superimposed.   
It was designed with the motivation of using a
theoretical model for extrapolation of ``low'' energy
accelerator and fixed-target data to the energies
necessary for the study of cosmic-ray interactions.   
SIBYLL agrees reasonably well with
available accelerator data with one notable
exception: the $\langle p_T\rangle$
distributions associated with particle production
in pp interactions are underestimated at large
Feynman x (x$_F > 0.15$) \cite{sibyll}.
The $\langle p_T\rangle$ distributions of parent
mesons are significant in studies like this because  
$\langle p_T\rangle$ is the dominant contributing factor to the lateral
distribution of underground muons.  At the Soudan 2
depth, the mean separation of muon pairs is comparable
to the size of the detector.
This means that the Monte Carlo simulation of muon
lateral spread is significant for a detector of our
size and depth since we need to determine correctly  
the rate of ``observed'' events of a given multiplicity
from the total number of events at our depth.  We have
tested the effect that the SIBYLL underestimation of 
$\langle p_T\rangle$ has on our analysis and discuss
this in Sec.\ \ref{sec:analysis}.
\par The nucleus-air interaction simulation is also
a modular component of the HEMAS cascade code. We
have used the NUCLIB \cite{nuclib} nuclear interaction
code for this stage.  In addition, the HEMAS cascade 
simulation package has been modified from its original 
form to include the 
effect of the local geomagnetic field, which has 
a strength of 0.59 Gauss and a magnetic
inclination and declination of $75.1^o$N and 
$0.85^o$E respectively at the Soudan 2 site.  Even though the
earth's magnetic field is weak, it plays a noticeable role
in determining the transverse displacement of the
underground muons because the distance traveled
by the cosmic ray muons through the atmosphere is very long.
Because the magnetic field near the Soudan site is nearly 
vertical, particles at large zenith angles are more greatly
affected than those near vertical.  
We have found that the addition
of the magnetic field to the cascade simulation has 
a negligible effect for events with zenith angle $\theta=15^o$.  
However, at $\theta=60^o$, the earth's magnetic field increases the 
mean transverse displacement of muons from the event core
at the Soudan 2 depth from 13 m with no magnetic field to 
18 m with magnetic field.
\par To determine the absolute single and multiple muon
rates in the detector, it is essential to understand the
composition and thickness of the rock overburden, as well
as to have a means to simulate the passage of the muons
through the rock.  A digitized map from
the US Geological Survey was used to determine the depth of rock
around the Soudan site.
The density of the rock at the Soudan site is
also variable.  Soudan is located in the ``Iron Range''
of northern Minnesota and the rock overburden is of non-uniform
composition with pockets of iron ore interspersed among
an overburden consisting mostly of Greenstone \cite{keith}.
To determine the density of the overburden, a fit to ``world
survey'' muon data \cite{crouch} was performed using 
a large body of muon data spanning the periods June, 1991
to March, 1996.  The effective rock density was determined
in 337 angular bins covering the angular region considered
in this analysis \cite{thesis}.
\par  The GEANT Detector Description Simulation \cite{geant} package developed by CERN
was used to propagate muons underground.  Muon
energy loss mechanisms are fairly well understood \cite{lohmann}, and in 
fact have been experimentally verified up to energies of 
$\sim 1$ TeV for muon energy loss in iron \cite{ccfr,mutron}.  
For Soudan 2 data this is the critical region of energy, since it 
defines the shape of the rise of the transmission probability
curve. (The transmission probability curve is the probability of a muon to
successfully reach the Soudan 2 level versus muon surface energy.)
We have compared GEANT to the available muon experimental
energy loss data with good results \cite{trost}.
\par Finally, the detector was simulated using a simplified
model which compares well against a more realistic
simulation of the response of the tracking calorimeter modules
and their readout electronics.
All cuts applied to the data were also applied to the Monte Carlo 
simulated events.

\section{Analysis}\label{sec:analysis} 
\subsection{Composition Models}
\par The energy and mass of the primary cosmic rays cannot be
determined on an event-by-event basis through indirect experimental methods
such as underground muon rates because of large fluctuations
in the atmospheric development of a cosmic ray cascade.  Instead, the
experimental approach is to assume a model of primary composition
at the top of the atmosphere as a function of mass and energy, 
and to use the Monte Carlo simulation to predict from the assumed 
composition the muon rate underground.  For simplicity, primary 
composition models generally divide primary cosmic rays into five
mass groups centered around the principal nuclei H, He, CNO, Ne-S, and Fe.
To narrow the field of possible test composition models, we have
defined three composition models which satisfy the
following criteria:
\begin{itemize} 
\item The model is theoretically motivated by an astrophysical model.
\item The model fits the available satellite and balloon
      direct measurement data in the low energy region ($<1000$ TeV/nucleus)
      in which this data is available. 
\item The model normalizes to the air-shower determined all-particle
      spectrum in the knee energy region.
\end{itemize}
\par In recent years the amount of direct measurement cosmic ray composition data, 
both satellite and balloon, has grown considerably.  Direct 
measurements currently extend up to almost 1000 TeV/nucleus \cite{sokol,jacee}.
Monte Carlo simulations show that Soudan 2  
should be sensitive to muon events generated by primary
cosmic rays in the energy region 
5 - 50000 TeV/nucleus.  Therefore, there is significant overlap 
between direct measurements and the energy region under study in our 
analysis.  
\par
\begin{figure}
\centerline{\psfig{figure=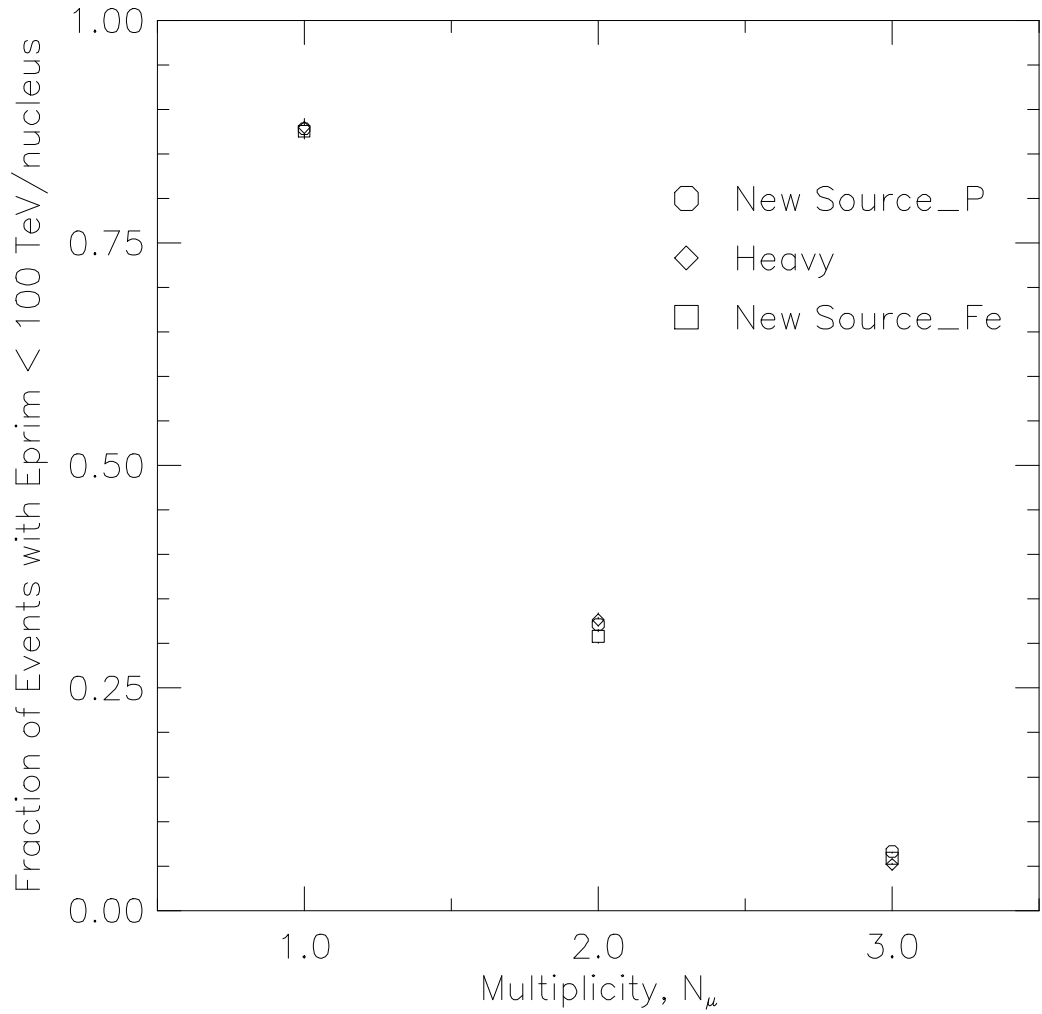,height=6.in,bbllx=0bp,bblly=100bp,bburx=600bp,bbury=650bp,clip=}}
\caption[]{\label{fig:frac}
The fraction of events at each multiplicity generated by  
primaries with energy less than 100 TeV/nucleus.  
}
\end{figure}

 Use of the available direct
measurement data performs two important functions in our 
analysis.   First, it tightly constrains the possible test composition
models, e.g. a composition model of pure protons over the 
entire energy range under study would be in obvious contradiction to the 
mixed composition observed in the lower energy regime by direct
measurements.  Secondly, proper normalization
to the direct measurement data allows for a test of the 
atmospheric cascade simulation at low energy.
To illustrate, Fig.~\ref{fig:frac} shows the fraction of events which are
produced by primaries of energy less than 100 TeV/nucleus 
as a function of multiplicity.  It is clear from the figure  
that greater than $85\%$ of single muon events are generated by
primaries in an energy region in which the composition is
well known.  Therefore, the absolute single muon rate in Soudan 2 can 
be used as an important test of the Monte Carlo simulation.
\par
\begin{figure}
\centerline{\psfig{figure=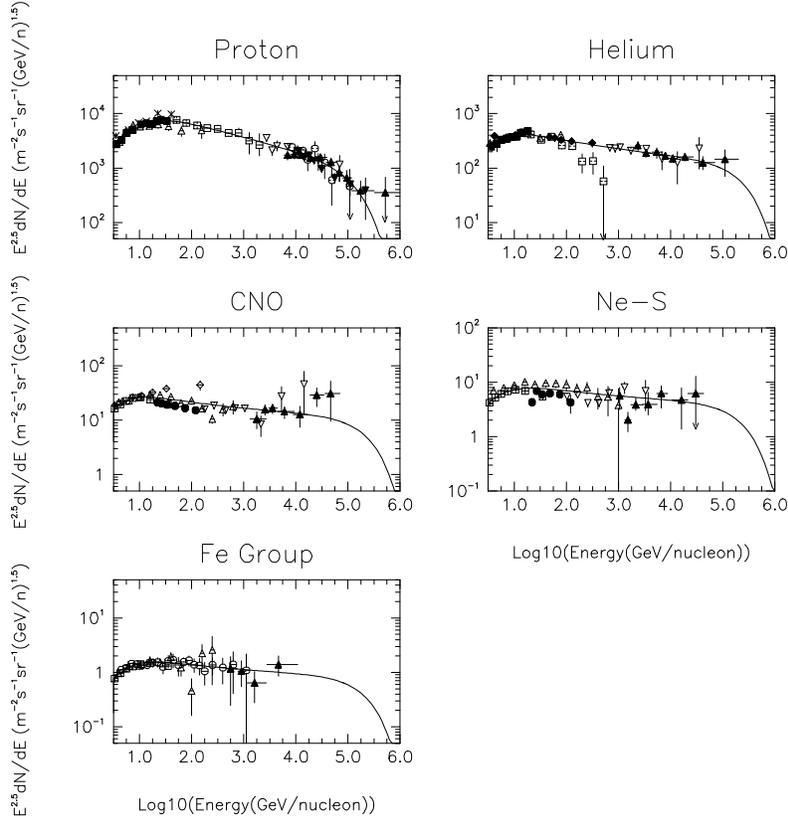,height=4.5in,bbllx=0bp,bblly=100bp,bburx=600bp,bbury=650bp,clip=}}
\caption[]{\label{fig:lndata}
Differential fluxes of the five mass groups used in the low energy
component of the New Source$\underline{ \ }$P and New Source$\underline{ \ }$Fe 
composition models.  
Muon events observed in Soudan 2 are products 
of primaries with energies above $\sim 10^3$~GeV/nucleon; however, it
is useful to show differential flux measurements at energies below
this to display trends in the data.
Below $\sim$10 GeV/nucleon, the observed flux of cosmic rays
is greatly affected by solar modulation, and the amplitude
of the data will depend upon the intensity of solar modulation at the time
the measurement was taken.  The fits for each of the five
mass groups, shown as solid curves, have 
been ``modulated'' at low energy in this figure by a 
factor of the form (1+$\alpha$E$^{-\beta})^{-1}$  
to fit the data, however this does not affect the low energy 
parametrizations given in the text for the energy region $>10^3$ GeV/nucleon 
which is of interest to us.
The direct measurement data have been compiled from the following sources: $\circ$\cite{ichi};
$\bullet$\cite{jul}; $\Box$\cite{ryan} for proton and helium and \cite{heao} for
heavier elements; filled $\Box$\cite{mass};
$\triangle$\cite{smith} for proton and helium and \cite{simon} for heavier elements; filled 
$\triangle$\cite{jacee}; $\bigtriangledown$\cite{sokol};
filled $\bigtriangledown$\cite{mubee}; $\Diamond$\cite{orth}; filled $\Diamond$\cite{rich};
$\ast$\cite{rosen}; and $\times$\cite{ormes}.  The ``Fe Group''
consists of the elements Z=26-28 for $\circ$ and $\Box$; Z=26 for filled $\triangle$; and
Z=26-30 for $\triangle$.}  
\end{figure}

\begin{figure}
\centerline{\psfig{figure=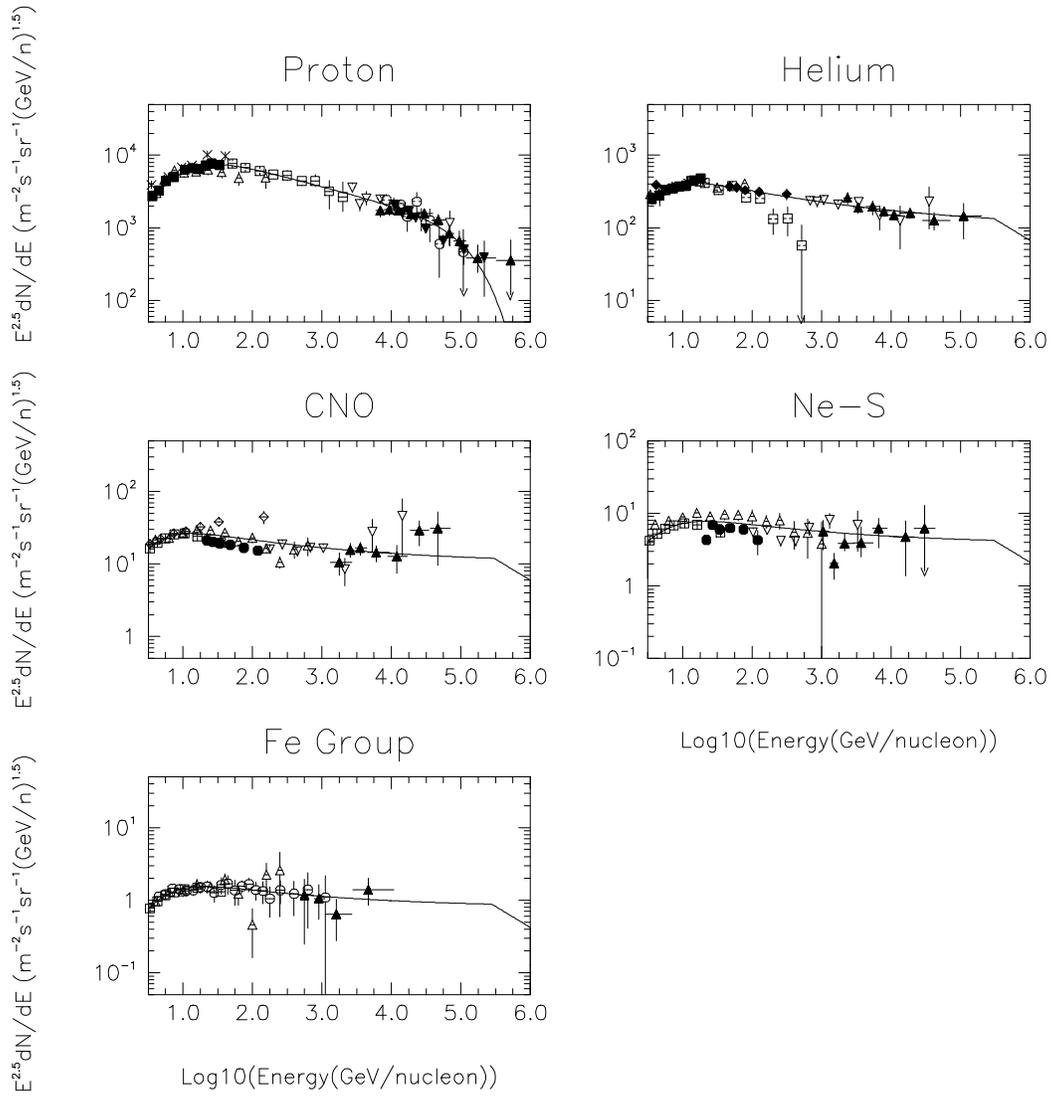,height=6.in,bbllx=0bp,bblly=100bp,bburx=600bp,bbury=650bp,clip=}}
\caption[]{\label{fig:hndata}
Differential fluxes of the five mass groups used in the Heavy composition model.  
The references for the direct measurement data can be found in the caption
for Figure~\ref{fig:lndata}.}
\end{figure}

 As has been pointed out elsewhere \cite{gaisser}, the availability
of new high energy direct measurement composition data has made some 
popular composition models somewhat obsolete.
In particular, we note that the versions of a Light test composition
model used by NUSEX \cite{nusex} and MACRO \cite{macro} no longer give good 
agreement to the available direct measurement data over the entire relevant 
energy range.                           
To formulate test composition models, we follow the lead of
Silberberg, {\it et al.} \cite{silb} and Stanev, Biermann, and Gaisser \cite{bier4} 
and assume a two-component model at low-energies, such that
the differential spectrum is described by
\begin{equation}
\label{eqn:comp2}
   \frac{dN}{dE} = K_1 E ^{-\gamma_1} + K_2 E ^{-\gamma_2}
\end{equation}                              
for each of the five mass groups.
A two component model follows naturally
from the assumption that the low energy cosmic rays come primarily
from two sources: supernova blasts into a homogeneous interstellar
medium(e.g. \cite{bland}) and supernova blasts into a stellar wind 
environment \cite{volk}. 
The latter is theorized to play a significant role for the heavy elements
and to produce a flatter spectrum than that of the former \cite{volk,silb}.  
To fit Eq.\ (\ref{eqn:comp2}) to the direct measurement data,
we have compiled data from a large number of direct measurement experiments.
These measurements are shown in Figs.~\ref{fig:lndata} and ~\ref{fig:hndata}.
We found that good agreement between Eq.\ (\ref{eqn:comp2}) and all
five mass groups could be obtained by using $\gamma_1$=2.75 and
$\gamma_2$=2.50, and by allowing the coefficients
K$_1$ and K$_2$ to have the values tabulated
in Table \ref{tab:lemod}.  
Eq.\ (\ref{eqn:comp2}), along with the values in Table \ref{tab:lemod},
is used to describe the low-energy component in all models considered in
our analysis.
The agreement of this low-energy component with the direct measurement data 
is shown in Figs.~\ref{fig:lndata} and ~\ref{fig:hndata}.
\par We then define three test composition models:
``New Source$\underline{ \ }$P'',``Heavy'', and ``New Source$\underline{ \ }$Fe''.  
The New Source$\underline{ \ }$P model, motivated by the model of 
Fichtel and Linsley \cite{linsley}, assumes that a new source consisting
entirely of protons predominates at energies around the knee.  
This type of proton-rich source is 
compatible with the ideas of the AGN particle acceleration model of Szabo 
and Protheroe \cite{agn4}.
In the New Source$\underline{ \ }$P model, we have assumed that the 
low-energy components have an exponential cutoff (as in \cite{bier4}),
such that the differential spectrum for these components becomes
\begin{equation} 
   \frac{dN}{dE}=(K_1 E ^{-\gamma_1} + K_2 E ^{-\gamma_2})e^{-\frac{E}{E_{cut}}}.
\end{equation}
The exponential cutoff energy, E$_{cut}$, is 
determined for the proton component by the direct measurement data.
The exponential cutoff for the heavier elements is allowed to extend out 
to higher energies, as suggested by reference \cite{volk} for a stellar
wind component, and as required by the helium data.
This cutoff for the heavier elements is also taken to be rigidity dependent.
The low-energy cutoff for all five mass groups are given
in Table \ref{tab:nmodel} and are shown in Fig.~\ref{fig:lndata}.
\par
\begin{figure}             
\centerline{\psfig{figure=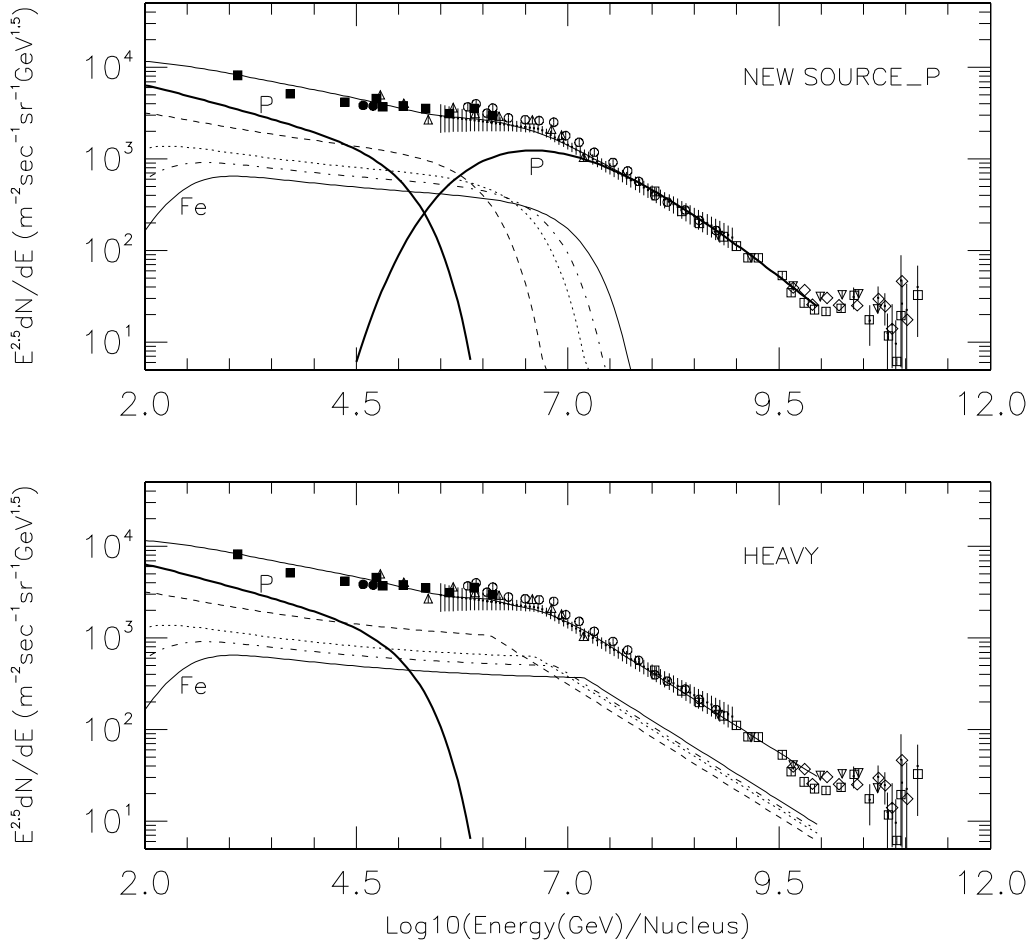,height=6.in,bbllx=0bp,bblly=100bp,bburx=600bp,bbury=650bp,clip=}}
\caption[]{\label{fig:allp}
The New Source$\underline{ \ }$P and Heavy composition models as compared to the all-particle spectrum.
The all-particle spectra of the two composition models have been normalized to
the Akeno parametrized all-particle spectrum \cite{akeno} (hashed region) in the knee region.  The rest of the
all-particle data are from the compilation by Stanev \cite{stanev}.  The five mass
groups shown are for P (bold), He (dash), CNO (dot), Ne-S (dot-dash), and Fe (solid).
The New Source$\underline{ \ }$Fe model is the same as 
New Source$\underline{ \ }$P
with the high-energy proton component replaced by iron.
}
\end{figure}

 The New Source$\underline{ \ }$P high-energy component, shown
in Fig.~\ref{fig:allp}, has the functional form
\begin{equation}                            
\label{eqn:hec}                           
   \frac{dN}{dE}=K_oe^{-A_oE^{-B_o}}E^{-\gamma_o},
\end{equation}    
with parameters for this component given 
in Table \ref{tab:hemod}.
We have chosen to normalize the high-energy component such that
the summed mass components of the model equal the 
all-particle spectrum as determined by Akeno \cite{akeno}.  
This choice of normalization and its effect on the underground muon 
rates will be discussed in the next section.
\par The second test model, the Heavy model, is motivated by the theoretical
model of Biermann, {\it et al.} \cite{bier1,bier4}, in which the stellar wind 
component extends out to energies up to $\sim Z\times 10^8$ GeV/nucleus.     
In the Heavy test model used here, each of the heavy mass groups (He,CNO,Ne-S,Fe) 
has the low energy form given in Eq.\ (\ref{eqn:comp2}) up
to a sharp break at energy E$_{knee}$, above which the mass groups follow
the single power-law
\begin{equation} 
   \frac{dN}{dE} \propto E ^{-\gamma_3}. 
\end{equation}
The E$_{knee}$ and $\gamma_3$ parameters for this model are given in 
Table \ref{tab:hmodel} and the agreement of this model with  
direct measurements is shown in Fig.~\ref{fig:hndata}.  The normalization 
to the all-particle spectrum closely matches that of 
New Source$\underline{ \ }$P and is shown in Fig.~\ref{fig:allp}.
\par The third test model, New Source$\underline{ \ }$Fe,
is exactly the same
as New Source$\underline{ \ }$P except that the high
energy protons have been replaced by iron to
test the opposite extreme.  
\par
\begin{figure}
\centerline{\psfig{figure=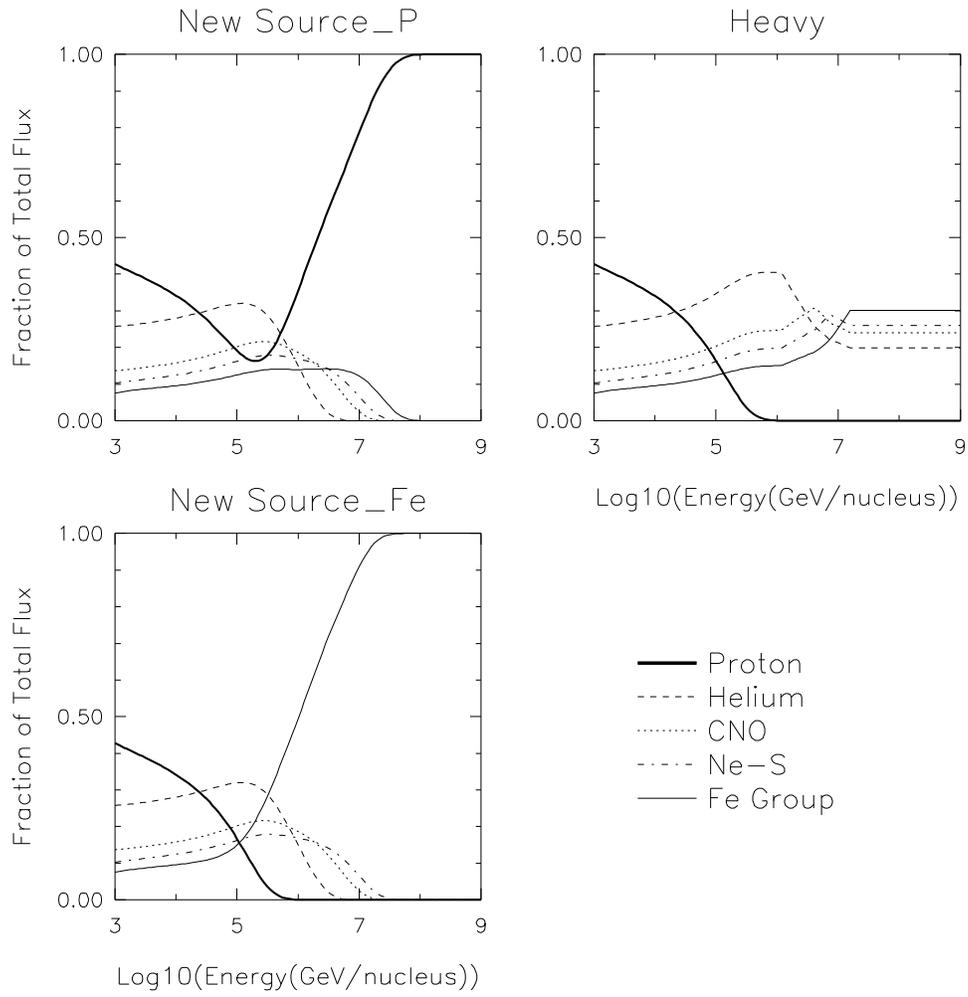,height=6.in,bbllx=0bp,bblly=100bp,bburx=600bp,bbury=650bp,clip=}}
\caption[]{\label{fig:rela}
The fractional composition of the three test composition models as a function
of energy.
}
\end{figure}

 The fractional composition 
for all three models is shown in Fig.~\ref{fig:rela}.
We note that even though a pure proton high energy component
is assumed for the New Source$\underline{ \ }$P model, this model is still
fairly heavy in what will be the critical energy region
for this study: $10^6-10^7$ GeV/nucleus.  In particular, we note
that the New Source$\underline{ \ }$P model is heavier than the MACRO 
Light \cite{macro} model in this energy region.
\subsection{Results}\label{sec:results}                  
\begin{figure}
\centerline{\psfig{figure=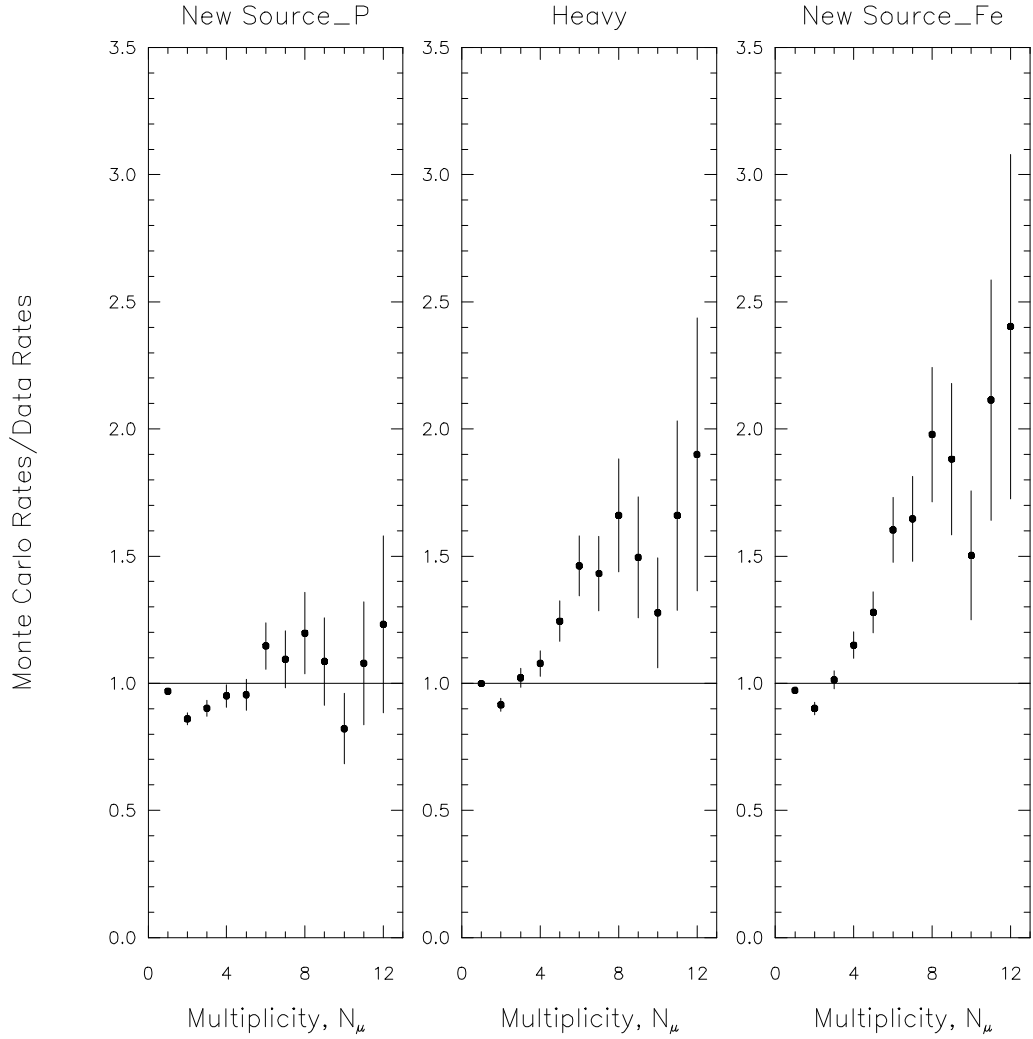,height=6.in,bbllx=0bp,bblly=100bp,bburx=600bp,bbury=650bp,clip=}}
\caption[]{\label{fig:rate}
Ratio of simulated to data absolute event rates as a function of event multiplicity.   
The ratio errors are calculated using the systematic and statistical
errors of the Data rates, and statistical errors only of the Monte Carlo rates.  See the
text for a discussion of Monte Carlo systematic errors. 
}
\end{figure}

\begin{figure}
\centerline{\psfig{figure=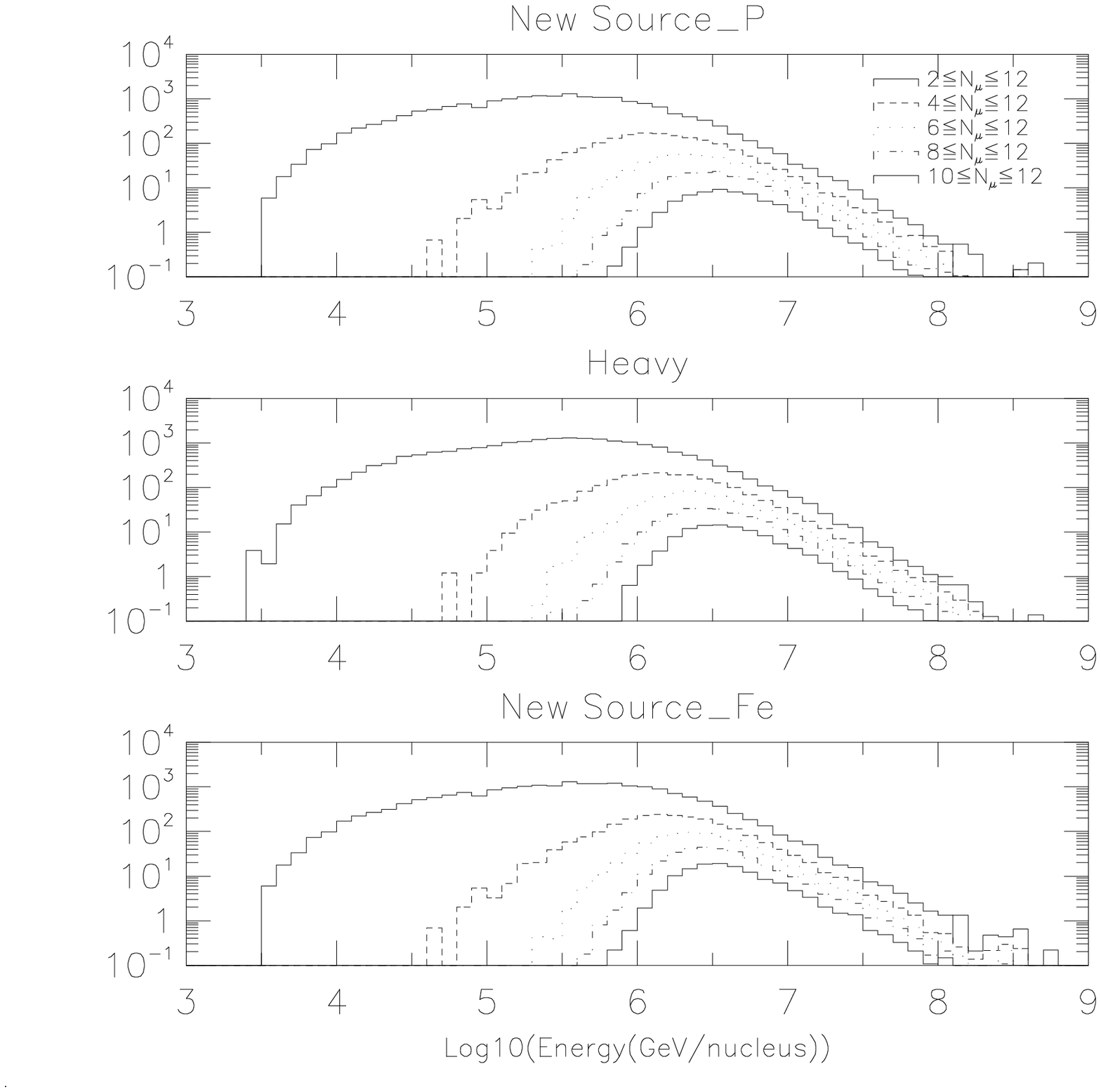,height=6.in,bbllx=0bp,bblly=100bp,bburx=600bp,bbury=650bp,clip=}}
\caption[]{\label{fig:energy}
Distributions of Monte Carlo multiple muon events as a function of primary energy.  The distributions
have been normalized to the total live time of the analysis.
}
\end{figure}

The results of this analysis are shown in Fig.~\ref{fig:rate}.
This figure shows the ratio of simulated multiple muon
event rates to observed event rates for each of the three test
composition models.
Fig.~\ref{fig:energy} shows the primary energy range to which
each multiplicity is sensitive.  We note that 90\% of the high multiplicity ($\geq 6$)
events come from a fairly well defined region of the energy
spectrum which is $8\times 10^5-1.3\times10^7$ GeV/nucleus.  This is the
energy region just below and around the knee of the energy spectrum.
From Fig.~\ref{fig:rate}, therefore, it is clear that the high multiplicity 
events offer a statistically significant test of  
composition of cosmic rays in this part of the all-particle spectrum.    
\par The agreement of the measured and simulated event
rates of each test composition model is tabulated in Table \ref{tab:chisq}.
$\chi^2$/d.o.f. in each case is calculated by comparing the absolute
event rates of the simulated and measured data for the seven multiplicities
6 through 12 which span the primary energy region of interest around
the knee of the all-particle spectrum. 
The data event rates used in this calculation include statistical and systematic errors, while
the simulated event rates include statistical errors only.                
Of the three trial composition models considered here, 
the New Source$\underline{ \ }$P simulated event rates clearly give 
the best agreement with the data in the knee energy region of 
the all-particle spectrum.
\par The observed absolute single muon rate agrees with the Monte Carlo
rates obtained with all three composition models to within 5\%.  As
already mentioned, $\sim85\%$ of single muon events come from a region
of the spectrum in which the composition is known from direct
measurements; therefore the agreement of the single muon observed
and Monte Carlo rates shows that the Monte Carlo is successful
at simulating muon events at low energy.
Of course, as we extend the measurement to higher energy, the
systematic uncertainties become more complex.  Here we discuss
two of the more important contributors to the systematic
uncertainty of the Monte Carlo simulation at energies around
the knee of the all-particle spectrum.
\subsubsection{All-particle spectrum normalization}  As already
mentioned, we have chosen to normalize the test composition models
to the Akeno determined all-particle spectrum \cite{akeno} in the
knee energy region.  The choice of normalization in this energy
range has some arbitrariness associated with it since the all-particle spectrum is
determined indirectly through air-shower measurements.  This
choice obviously has an effect on the high multiplicity underground muon rates.
The Akeno group \cite{akeno} established a range in which the
all-particle spectrum might fall based on their own measurements
using various techniques and the measurements of other experiments.
Their parametrized form of their measured all-particle spectrum,
used in the analysis presented here, lies along the low end of this range.  If we were to 
increase the normalization of the test
composition models towards the upper end of their range, the effect
would be to push the results further {\it away} from the heavy composition
models.  
\par  We also compare the normalization of the test model all-particle
spectra to the recently reported result of the Tibet AS$\gamma$ \cite{tibet} air
shower collaboration.  The Tibet air-shower array operates at the
high altitude of 4300 meters above sea level, corresponding to
an atmospheric depth of 606 g/cm$^2$.  This is an ideal altitude for
studying the all-particle spectrum in the energy region just before and around the knee,
since air showers generated by primaries
in the energy range $\sim 10^5-10^8$ GeV reach their maximum shower
development close to the altitude of the experiment, and hence their
fluctuations at the point of sampling are at a minimum.  Also, the
average shower sizes generated by primaries of different atomic mass
become nearly identical at sampling heights near shower maximum.  This
is very important for eliminating the systematic dependence of the measurement
of primary energy on the assumed primary composition.  
\par The all-particle spectra of the test composition models considered in
this analysis lie about $22\%$ lower than the Tibet measurements at energies
of $10^{5.75}$ GeV, and about $9\%$ higher at energies of $10^7$ GeV.  A
renormalization of the all-particle spectrum of the Heavy model to the
Tibet spectrum yields underground high-multiplicity event rates which
are $\sim 10\%$ higher than the present rates, again pushing the Heavy model
further away from the data.
\subsubsection{Hadronic interaction model} 
As already noted, one of the uncertainties in the hadronic interaction
model is that SIBYLL tends to underestimate the increase in
$\langle p_T\rangle$ with x$_F$ (x$_F>0.15$) for particle production
in pp interactions.
On the other hand, an older
version of a Monte Carlo hadronic interaction code, HEMAS \cite{hemas}, 
produces a significantly greater mean transverse momentum in the 
forward fragmentation region than that of SIBYLL, and is
in fact likely to overestimate the $\langle p_T\rangle$ at large x$_F$ \cite{sibyll}.
We have run a test using the Heavy model as input and SIBYLL as the main driving code, but scaling
the transverse momentum in each SIBYLL hadronic interaction according to
the HEMAS $\langle p_T\rangle$ distributions.  In this way, we were able to extract
and test the effect that just this one aspect of the hadronic interaction
code has on the Monte Carlo simulated Heavy model rates.  The result is that approximately 
$10\%$ fewer Monte Carlo events with multiplicities $\geq 6$ are seen in              
the detector given the larger $\langle p_T\rangle$ distributions of the HEMAS 
hadronic interaction code.  This lowers the
Heavy composition model rates in the direction of the data, resulting
in an improved $\chi^2$/d.o.f. agreement between the data and the Heavy
model.  The $\chi^2$/d.o.f.  
for the $\langle p_T\rangle$ scaled Heavy model rates (calculated,
as before, by comparing the absolute event rates of the simulated and
measured data for the seven multiplicities 6 through 12) is 55/7 (C.L. = 1.8$\times 10^{-7}$\%).    

\section{Conclusion}\label{sec:conclude}
\par  We have shown that the multiple muon rates observed           
in the Soudan 2 detector are sensitive to the nuclear composition
of cosmic rays in the energy region $8\times 10^5-1.3\times10^7$ GeV/nucleus,
which is the energy region just before and around the knee in the 
cosmic ray all-particle spectrum.
The composition model favored in this work includes
an enhanced component of protons in this energy region.
This component is compatible with the AGN model of particle
acceleration of Szabo and Protheroe \cite{agn4}, which shows
a contribution of protons due to AGN sources localized 
in the knee energy region of the spectrum.      
\par
\begin{figure}
\centerline{\psfig{figure=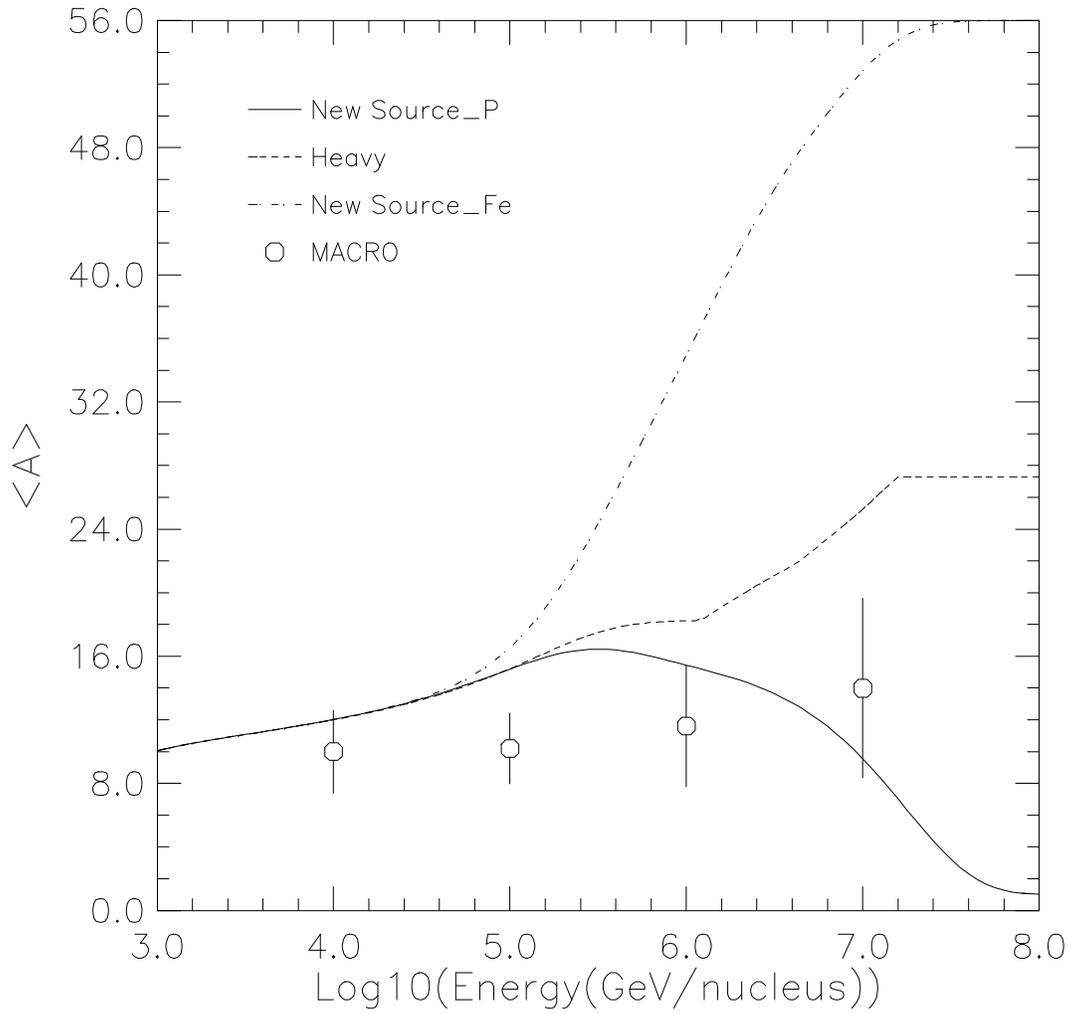,height=6.in,bbllx=0bp,bblly=100bp,bburx=600bp,bbury=650bp,clip=}}
\caption[]{\label{fig:macroa}
Average atomic mass versus primary energy for each of the three test composition models, and
as determined by MACRO \cite{macro2}.
}
\end{figure}

 The light composition model favored in this analysis is compatible with
previous results from Soudan 1 \cite{urmi} and Soudan 2 \cite{nat}, 
in which a small surface array
was operated in conjunction with the underground detector. 
The result is also in agreement with the recent multiple
muon analysis performed by MACRO \cite{macro2}, as is shown in Fig.~\ref{fig:macroa}.
Results from Fly's Eye \cite{bug} favor a  
mixed to heavy composition at energies near $3\times10^8$ GeV/nucleus. 
We note that the Fly's Eye result does not directly contradict
our result, since our measurement applies to lower primary energies.
\section*{Acknowledgements}
\par We are very grateful to Tom Gaisser for making the HEMAS/SIBYLL cascade 
simulation package available to us, and for providing
valuable advice.
\par This work was supported by the U.S. Department of Energy, 
the U.K. Particle Physics and Astronomy Research Council,
and the State and University of Minnesota.  We wish to thank
the staffs of the collaborating institutions, the Minnesota
Department of Natural Resources for use of the facilities
at the Soudan Underground Mine State Park, the staff of the
park for their support (particularly Park Managers D. Logan and
P. Wannarka), and the Soudan 2 mine crew: B. Anderson, J. Beaty, G. Benson,
D. Carlson, J. Eininger and J. Meier.

% now the references. delete or change fake bibitem. delete next three
%   lines and directly read in your .bbl file if you use bibtex.
%\bibliography{mymmu} for use with bibtex, this is impossible to work with!

% \begin{figure}
% \caption{}
% \label{}
% \end{figure}

                         
%\begin{table}
%\caption{}
%\label{}
% \end{tabular}
% \end{table}

\begin{table}                           
\caption{Efficiency and background for reconstructed single muon events.}
\label{tab:eff}
\begin{tabular} {l c c c c c}
Run  & Date & Total Scanned & Muon Events & Efficiency & Background\\  
     &      &    Events     &  Passing Cuts &          &       \\
\hline
29052 &  7/21/91 &  1000 & 273 & $\frac{271}{273}=0.993 \pm 0.005 $ & $\frac{4}{275}=0.015 \pm 0.007  $ \\
31111 & 10/14/91 &   900 & 321 & $\frac{318}{321}=0.991 \pm 0.005 $ & $\frac{8}{326}=0.025 \pm 0.009  $ \\
\hline                                         
Total & & 1900 & 594 & $\frac{589}{594} = 0.992\pm 0.004$ & $\frac{12}{601}=0.020\pm 0.006$\\
\end{tabular}
\end{table}

\begin{table}                           
\caption{Efficiency of the software at finding multiple muon 
         events satisfying the cuts specified in the text.}
\label{tab:mmueff}
\begin{tabular} {l c c c c }
Run  & Date & Total Events & Multiple Muon Events & Efficiency \\  
\hline
29052 &  7/21/91 &  2070 & 23 & 1.\\
31111 & 10/14/91 &   900 &  7 & 1.\\ 
28132 &  6/13/91 &  1858 & 12 & 1.\\
33009 &  1/01/92 &  1868 & 13 & 1.\\
35018 &  3/31/92 &  1795 & 19 & 1.\\
35520 &  4/26/92 &  1428 & 16 & 1.\\
35848 &  5/12/92 &  1881 & 19 & 1.\\
\hline                                   
Total & & 11800 & 109 & $\frac{109}{109} = 1.$\\
\end{tabular}
\end{table}

\begin{table}
\caption{Number of events and event rate observed at each multiplicity.}
\label{tab:mult}                                  
\begin{tabular} {l c c c c }
Multi- &  Events & Events $\pm$ stat $\pm$ syst & Rate (hr$^{-1}$)$\pm$ stat $\pm$ syst\\  
plicity &(uncorrected) & (corrected) & (corrected) \\
\hline                                 
1  &  724792 & 716024 $\pm$ 846 $\pm$ 5249 & 530.87 $\pm$ 0.63  $\pm$ 3.89 \\              
2  &   17237 &  17237 $\pm$ 131 $^{+417}_{-0}$  & 12.780 $\pm$ 0.097 $^{+0.309}_{-0}$ \\
3  &    2813 &   2813 $\pm$  53 $^{+68}_{-0}$   & 2.086 $\pm$ 0.039 $^{+0.050}_{-0}$  \\
4  &     905 &    905 $\pm$  30 $^{+22}_{-0}$   & 0.671 $\pm$ 0.022 $^{+0.016}_{-0}$  \\
5  &     385 &    385 $\pm$  20 $^{+9.4}_{-0}$  & 0.285 $\pm$ 0.015 $^{+0.007}_{-0}$ \\
6  &     172 &    172 $\pm$  13 $^{+4.1}_{-0}$  & 0.128 $\pm$ 0.010 $^{+0.003}_{-0}$ \\
7  &     102 &    102 $\pm$  10 $^{+2.5}_{-0}$  & 0.0756 $\pm$ 0.0074 $^{+0.0019}_{-0}$\\
8  &      58 &     58 $\pm$  7.6 $^{+1.4}_{-0}$ & 0.0430 $\pm$ 0.0056 $^{+0.0010}_{-0}$ \\
9  &      42 &     42 $\pm$  6.5 $^{+1.0}_{-0}$ & 0.0311 $\pm$ 0.0048 $^{+0.0007}_{-0}$ \\
10 &      37 &     37 $\pm$  6.1 $^{+0.90}_{-0}$ & 0.0274 $\pm$ 0.0045 $^{+0.0007}_{-0}$ \\
11 &      20 &     20 $\pm$  4.5 $^{+0.48}_{-0}$ & 0.0148 $\pm$ 0.0033 $^{+0.0004}_{-0}$\\
12 &      13 &     13 $\pm$  3.6 $^{+0.31}_{-0}$ & 0.0096 $\pm$ 0.0027 $^{+0.0003}_{-0}$\\
\end{tabular}
\end{table}                                                               

\begin{table}                                          
\caption{Low-energy parameters of the New Source$\underline{ \ }$P, 
         New Source$\underline{ \ }$Fe, and Heavy  
         composition models.
         The units of K$_1,_2$ are 
         [m$^{-2}$s$^{-1}$sr$^{-1}$(GeV/nucleus)$^{\gamma_1,_2 - 1}$].}
\label{tab:lemod}
\begin{tabular} {l c c c c c c }
Mass Group &  Zeff & Aeff  & K$_1$ & $\gamma_1$ & K$_2$ & $\gamma_2$\\  
\hline
H    &  1 &  1 &  20830.& 2.75 &   0. & 2.5 \\
He   &  2 &  4 &  7750. & 2.75 & 840. & 2.5 \\
CNO  &  7 & 14 &  3545. & 2.75 & 550. & 2.5 \\
Ne-S & 12 & 24 &  2655. & 2.75 & 445. & 2.5 \\
Fe   & 26 & 56 &  2120. & 2.75 & 335. & 2.5 \\
\end{tabular}
\end{table}

\begin{table}
\caption{The low-energy component energy cutoff used in 
         the New Source$\underline{ \ }$P and 
         New Source$\underline{ \ }$Fe 
         composition models.  
         E$_{cut}$ is given in (GeV/nucleus).
}
\label{tab:nmodel}
\begin{tabular} { l c }
Mass Group & E$_{cut}$\\  
\hline
H & $1.5\times 10^5$ \\
He & $1.0\times 10^6$ \\
CNO & $3.5\times 10^6$ \\
Ne-S & $6.0\times 10^6$ \\
Fe & $1.3\times 10^7$ \\
\end{tabular}
\end{table}

\begin{table}
\caption{High-energy parameters of the 
          New Source$\underline{ \ }$P and 
         New Source$\underline{ \ }$Fe 
         composition model.  The units of K$_o$ 
         are [m$^{-2}$s$^{-1}$sr$^{-1}$(GeV/nucleus)$^{\gamma_o - 1}$]
         and A$_o$ is given in (GeV/nucleus)$^{B_o}$.
}                                                               
\label{tab:hemod}
\begin{tabular} { l c c c }
K$_o$ & A$_o$ & B$_o$ & $\gamma_o$ \\  
\hline
2.75E9 & 327. &  0.322 & 3.3 \\
\end{tabular}
\end{table}

\begin{table}
\caption{High-energy parameters of the Heavy composition model. 
         E$_{knee}$ is given in GeV/nucleus.}
\label{tab:hmodel}
\begin{tabular} { l c c }
Mass Group & E$_{knee}$ & $\gamma_3$ \\
\hline
He   & $1.2\times 10^6$ & 3.08 \\
CNO  & $4.2\times 10^6$ & 3.08 \\
Ne-S & $7.2\times 10^6$ & 3.08 \\
Fe   & $15.6\times 10^6$ & 3.08 \\
\end{tabular}
\end{table}           
                             
\begin{table}
\caption{Calculated values of $\chi^2$, $\chi^2$ per degree of freedom (d.o.f.),      
    and confidence level for each of the test composition models.
    See text for further discussion.}
\label{tab:chisq}
\begin{tabular} { l c c c c }
Composition Model & $\chi^2$ & d.o.f. & $\chi^2$/d.o.f. & C.L. (\%)\\
\hline
New Source$\underline{ \ }$P& 8.7 & 7 & 1.2 & 28 \\
Heavy  & 103 & 7 & 15 & 2.1$\times 10^{-17}$\\     
New Source$\underline{ \ }$Fe & 236  & 7 & 34 & 3.0$\times 10^{-45}$ \\
\end{tabular}
\end{table}               
\end{document}